\documentclass[a4paper,aps,prd,showpacs,twocolumn,nofootinbib]{revtex4-2}
\usepackage[T1]{fontenc}
\usepackage[utf8]{inputenc}
\usepackage[brazil,english]{babel}
\usepackage{color}
\usepackage{amsmath,amssymb}
\usepackage{graphicx}
\usepackage[unicode,bookmarks=false,colorlinks=true,linkcolor=blue,urlcolor =blue, citecolor=blue, anchorcolor=blue]{hyperref}

\usepackage{appendix}

\begin{document}
 
\title{Rotating regular black holes and other compact objects with a Tolman type potential as a regular interior for the Kerr metric}

\author{ Angel D. Masa$^{1}$ and  Vilson T.
Zanchin$^{2}$}
 \affiliation{Departamento de F\'{i}sica, Facultad de Ciencias Exactas y Naturales,
Universidad de Cartagena, Carrera 50 $\#$ 24-120, 130015 -- Cartagena de Indias, Bol\'{i}var,
Colombia\vspace*{.15cm}\\
$^{2}$Centro de Ci\^{e}ncias Naturais e Humanas,
Universidade Federal do ABC,
 Avenida dos Estados
 5001, 09210-580 -- Santo Andr\'{e}, S\~{a}o Paulo, Brazil}

\begin{abstract}

We obtain a new class of stationary axisymmetric spacetimes by using the Gürses-Gürsey metric with an appropriate mass function in order to generate a rotating core  of matter that may be smoothly matched to the exterior Kerr metric. The same stationary spacetimes may be obtained by applying a slightly modified version of the Newman-Janis algorithm to a nonrotating spherically symmetric seed metric. The starting spherically symmetric configuration represents a nonisotropic de-Sitter type fluid whose radial pressure $p_r$ satisfies an state equation of the form $p_r=-\rho$, where the energy density $\rho$ is chosen to be the Tolman type-VII energy density [R. C. Tolman, Phys. Rev. {\bf 55}, 364 (1939)]. The resulting rotating metric is then smoothly matched to the exterior Kerr metric, and the main properties of the obtained geometries are investigated. All the solutions considered in the present study are regular in the sense they are free of curvature singularities.  Depending on the relative values of the total mass $m$ and rotation parameter $a$, the resulting stationary spacetimes represent different kinds of rotating compact objects such as regular black holes, extremal regular black holes, and regular starlike configurations.

\end{abstract}



\maketitle


\section{Introduction }

General Relativity theory developed into a powerful tool for studying the structure of massive and compact bodies and the  gravitational effects generated by such objects. However, the same tool results not sufficient when trying to explain a particular property of one of the most beautiful predictions of the theory, the black holes, more specifically the singularities associated to black holes.  
This pathology is predicted by the singularity theorems~\cite{Hawking:1973uf}, is present in the final state of collapsing massive matter distributions, and is inevitable if the matter satisfies some reasonable energy conditions. 
A simple way of characterizing a singularity is by means of the curvature scalars, whenever some of the curvature scalars diverge a gravitational singularity occurs.  
Curvature singularities are intrinsic features of the manifold in the sense that it is impossible to eliminate them by any change of coordinates. In order to get rid of this kind of divergences in blackhole solutions, several alternative approaches have been proposed, the result being the regular black holes. These are compact objects with horizons but free of singularities. We may say that the search for regular black holes started with the work by Bardeen~\cite{bardeen1968non}, and with the passing of time many other models were built, see Refs.~\cite{Frolov:1989pf,Dymnikova:1992ux,Beato1998,br01,Lemos:2011vz,Hayward:2005gi,BronFab2006,br12,Lemos:2017vz,Simpson:2019mud,Berry:2020ntz,Bronnikov:2005gm} for a small sample of regular blackhole models in general relativity, see also Ref.~\cite{Ansoldi:2008jw} for an interesting review and Ref.~\cite{AngEnVil2021} for more references on the subject. In these models, the usual way to avoid singularities is by assuming a kind of matter content that violates the energy conditions, such as perfect fluids with negative pressures, so that the singularity theorems do not apply.

Most of the regular blackhole models found in the literature, including the ones mentioned in the last paragraph, are restricted to nonrotating configurations. Even though they represent idealized configurations, many interesting properties of very compact objects in general relativity have been exploited in such studies. On the other hand, observational data show that astrophysical objects are spinning.  In this aspect, even though an analogous of the Birkhoff theorem for stationary spacetimes is missing, the Kerr geometry~\cite{Kerr:1963ud} is accepted as a suitable or the most realistic solution among the several blackhole solutions that can be used to describe collapsed matter with angular momentum.  
However, the Kerr metric brings with it certain complications such as a singularity in the form of a ring, causality violation, and closed timelike curves (see, e.g. Ref.~\cite{Carroll:2004st}). A simple solution to these issues is to replace the problematic region with a regular matter source in the same way as done for nonrotating regular blackhole configurations. This approach leads us to one of the most interesting problems in relativistic astrophysics, that is finding a metric which solves the Einstein equations, that can be used as interior geometry for a rotating compact object, and that can be smoothly matched to the exterior Kerr solution. In fact, many works tried to obtain an interior solution for the Kerr geometry (see, e.g. Refs. \cite{Hernandez:1967zza,Cohen1967NoteOT,DeLaCruz1968,HartleThorne1968,Florides1973ARS,Florides1975,Israelrot1970,Hamuty1976,Collas1976},  see also Refs.~\cite{Krasinski:1976vyc} and \cite{Majidi2017} for reviews), but several solutions reported are valid only in the slow rotation limit (see for instance Refs.~\cite{Hernandez:1967zza,Cohen1967NoteOT,DeLaCruz1968,HartleThorne1968,Florides1973ARS,Florides1975} and \cite{Uchikata:2014kwm}).  Other proposed models for rotating compact objects in which the interior solution is smoothly matched to the Kerr metric, i.e., without any further approximation and avoiding the presence of surface layers bearing any king of matter, are characterized by configurations presenting some nonphysical or unexpected properties for a rotating star, such as negative mass~\cite{Israelrot1970}, a singular behavior~\cite{Hamuty1976}, or require some exotic matter such as nonisotropic fluids with negative pressure profiles~\cite{Collas1976}. In fact, since the discovery of the Kerr metric, this has been a subject of continuous interest in the literature, see Refs. \cite{Hernandez:2017,Herrera:2017,Simpson:2021dyo,Simpson:2021zfl} and references therein for some promising models investigated more recently.

In general, finding an exact solution to the Einstein equations that models a rotating compact object is an arduous and complicate task, and that is why many authors opt for numerical approaches. Nevertheless, there are several alternative methods to find new solutions describing stationary spacetimes \cite{Newman:1965tw,Demianski:1972uza,Gurses:1975vu,Krasinski:1976vyc,Clement:1998,Glass:2004,Viaggiu:2008,Ovalle:2021}. One of these alternative procedures is characterized by an appropriate complex coordinate transformation, an interesting trick revealed by Newman and Janis~\cite{Newman:1965tw} when obtaining the Kerr metric in Boyer-Lindquist coordinates from the Schwarzschild metric. This is the Newman-Janis algorithm (NJA), an approach that has been used also to derive the Kerr-Newman geometry from the static Reissner-Nordström metric~\cite{Newman:1965my}. 

Initially the NJA was shown to work well for the Kerr and Kerr-Newman vacuum solutions, but it was later tried to generate new solutions in the presence of rotating matter distributions. 
 For instance, in Ref.~\cite{Herrera1982} the NJA was applied to a generalized version of the interior Schwarzschild solution as an attempt to generate a regular source for the Kerr metric. In principle, the interior stationary metric could be joined smoothly to the Kerr metric as long as the arbitrary functions introduced satisfy suitable constraints. The resulting source is a nonisotropic rotating fluid,  but the authors conclude that the model is ``quite imperfect to be considered as the interior Kerr metric''. In fact, the straightforward application of the NJA to generic spherically symmetric static geometries leads to important issues, such as the facts that the resulting rotating metric cannot be transformed back to Boyer-Lindquist coordinates,
 and that the resulting energy-momentum tensor may present some nonphysical properties. To circumvent such problems, extensions of the original Newman-Janis procedure have been proposed. See Refs. ~\cite{Drake:1997hh,Drake:1998gf,Viaggiu:2006mx,Bambi:2013ufa,Azreg-Ainou:2014nra,Azreg-Ainou:2014aqa,Azreg-Ainou:2014pra} for some examples of  modifications of the NJA proposed in the literature, see also Ref.~\cite{Erbin:2016lzq} for a review. In this way, starting from a known static and spherically symmetric spacetime, these improved algorithms have been used successfully to generate rotating geometries.

In Ref.~\cite{Drake:1997hh} the NJA was adapted to be applied to any static spherically symmetric spacetime, and the smooth boundary conditions to join the  interior rotation metric to the exterior Kerr metric through an oblate spheroidal surface were explicitly formulated.
In Ref.~\cite{Viaggiu:2006mx} the NJA was applied to the interior Schwarzschild solution, as in Ref.~\cite{Herrera1982}, and the possible matter source for the geometry in the slowly rotating limit was considered. In the same work~\cite{Viaggiu:2006mx}, a second example of rotating interior metric for the Kerr metric was investigated by applying the NJA to the static spherically symmetric anisotropic fluid sphere found in Ref.~\cite{Stewart1982}.
In fact, this and other modified versions of the original Newman-Janis procedure, with a few different generalizations, have been employed successfully to obtain several solutions representing rotating regular black holes \cite{Modesto2010,Bambi:2013ufa,Azreg-Ainou:2014pra,Toshmatov:2014nya,Neves:2014aba,Ghosh:2015pba,Tsukamoto:2017fxq}. Similarly to the case of static regular blackhole models, these rotating regular black holes violate some of the energy conditions. It is worth mentioning that extensions of the NJA have also been used to obtain rotating black hole solutions in alternative theories of gravity~\cite{Caravelli:2010,Azreg-Ainou:2011tkx,Ghosh:2013,KumarGosh:2018}.

Inspired by the works mentioned in the last paragraphs, the chief goal of the present work is constructing a rotating interior solution that can be smoothly matched to the exterior Kerr metric in such a way that the resulting stationary axisymmetric spacetime is free of curvature singularities. Our strategy is to apply a modified version of the NJA to the Tolman type-VII potential~\cite{Tolman:1939}, which describes a well behaved static spherically symmetric matter distribution. The resulting stationary metric is the same obtained by following the well set approach by Gürses and Gürsey\cite{Gurses:1975vu}, once one starts with the same Tolman potential, i.e., with the same mass function.
The motivation for using the Tolman VII ansatz as the seed solution in the NJ procedure is that the metric is singularity-free, with the metric potentials being continuous and well-behaved functions inside of the compact object. The corresponding energy density presents a quadratic falloff profile, decreasing
monotonically from the center toward the edge of
the spherical distribution of matter, what represents a realistic physical model~\cite{Raghoonundun2015}.

 The present work is organized as follows. In Sec.~\ref{Sec:GGmetric} the Gürses-Gürsey metric is described in brief. In Sec.~\ref{Sec:RotTol} we present a new stationary metric that represents rotating spheroidal compact objects, and the smooth matching to the exterior Kerr metric is performed. Section~\ref{Sec:RotAniSph} is dedicated to study the main physical properties of the new rotating anisotropic spheroids. In this section we investigate the matter sources of the geometries, the curvature scalars and possible singularities, the horizons, the ergospheres, and the nature of the matching surfaces. In Sec.~\ref{Sec:Conclusion} we make final comments, and in Appendix \ref{appendixA} we expose some technical details regarding the smooth junction conditions between the inner matter metric and the outer Kerr metric not presented in the text.

\section{The Gürses and Gürsey metric}\label{Sec:GGmetric}

The present study makes used of the Gürses-Gürsey metric\cite{Gurses:1975vu}, which, in the Boyer-Lindquist coordinates  $ (t,\, r,\, \theta,\, \varphi)$, reads
\begin{equation}\label{eq:RST}
\begin{split}
    ds^{2}=&-\left(1-\dfrac{2\,r\,M(r)}{\Sigma}\right)dt^{2}+\dfrac{\Sigma}{\Delta(r)}dr^{2}+\Sigma\, d\theta^{2}\\ &
    -\dfrac{4\,r\,M(r)\,a\sin ^{2}\theta}{\Sigma(r,\theta)}dt\,d\phi \\
         & +\sin^{2}\theta\left(r^{2}+a^{2}+\dfrac{2\,r\,M(r)\,a^{2}{\rm sin}^{2}\theta}{\Sigma}\right) d\phi^{2} ,
\end{split}
\end{equation}
where $M(r)$ is an arbitrary function of the radial coordinate $r$ alone, and $\Delta(r)$ and $\Sigma(r,\theta)$ are given by 
\begin{eqnarray}
   &  &    \Delta(r)=r^{2}+a^{2}-2\,r\,M(r),  \label{eq:Delta} \\
  & & \Sigma(r,\theta)=r^{2}+a^{2}\,{\rm cos}^{2}\,\theta, \label{eq:Sigma}
\end{eqnarray}  
respectively. As shown in Ref.~\cite{Gurses:1975vu}, the stationary metric \eqref{eq:RST} is a solution of the Einstein equations that describes spinning systems, each different system of this kind being completely determined by the mass function $M(r)$.

Metric \eqref{eq:RST} may be obtained by means of the Gürses-Gürsey approach. The starting point is the Kerr-Schild metric $ds^2 = \eta_{\mu\nu}+ 2H\, k_\mu k_\nu$, 
where $\eta_{\mu\nu}$ is the Minkowski metric, $H$ is an arbitrary function, and $k_\mu$ is a vector field tangent to the Kerr principal lightlike congruence. A complex coordinate transformation brings the original metric into another. In particular, one may start with a static, spherically symmetric spacetime metric in the form 
\begin{equation}\label{eq:Smetric}
    ds^{2}=-f(r)dt^{2}+f^{-1}(r)dr^{2}+r^{2}\left(d\theta^2 + \sin^2\theta\, d\varphi^2\right), 
\end{equation}
where  $(t,\, r,\, \theta,\, \varphi)$ are Schwarzshild-type coordinates and $f(r)$ is a function of the the areal radius $r$ only, which is usually written as
\begin{equation}
    f(r) = 1- \dfrac{2M(r)}{r}. \label{eq:fstatic}
\end{equation}
In such a case, the Gürses-Gürsey approach transforms the static metric \eqref{eq:Smetric} into the Kerr-Schild form in which the function $H$ is given by $ \frac{r\, M(r)}{\Sigma}$. At the end, by means of a standard coordinate transformation, the resulting metric may be put into the form \eqref{eq:RST}, which describes a stationary spacetime. 

The Gürses and Gürsey rotating metric \eqref{eq:RST} and its generalizations, including a conformal factor, have been derived for the first time by the non-complexification procedure of the NJA in Refs.~\cite{Azreg-Ainou:2014nra,Azreg-Ainou:2014aqa,Azreg-Ainou:2014pra}.
 For more details on the NJA and its generalizations see, e.g., Refs.~\cite{Drake:1997hh,Drake:1998gf,Viaggiu:2006mx,Bambi:2013ufa,Azreg-Ainou:2014nra,Azreg-Ainou:2014aqa,Azreg-Ainou:2014pra}.

 As also well known, metric \eqref{eq:RST} may be obtained by means of the NJA~\cite{Newman:1965tw}. Such an algorithm starts with the same static spherically symmetric metric and after a complexification process, by properly taking the real parts of the relevant quantities, function $f(r)$ transforms as $  f(r)\to F(r,\,\theta)=1-\frac{2\, r\,M(r)}{\Sigma}$, with the mass function $M(r)$ keeping its original form. As in the approach by Gürses and Gürsey, the complete transformation leads the metric \eqref{eq:Smetric} into the stationary metric \eqref{eq:RST}.

\section{A new rotating solution}\label{Sec:RotTol}

\subsection{The  interior solution}\label{Sec:InterSol}

Inspired in the work by Tolman~\cite{Tolman:1939}, 
we consider the metric potential $f(r)$ of the static spherically symmetric spacetime~\eqref{eq:Smetric} in the form 
\begin{equation}\label{eq:Tpotential}
    f(r)=1-\dfrac{r^2}{R^2}+\dfrac{r^4}{A^4},\quad    
\end{equation}
where $R$ and $A$ are two arbitrary constant parameters. 
This is the Tolman metric potential associated to his type-VII solution.

The metric \eqref{eq:Smetric}, with $f(r)$ given by \eqref{eq:Tpotential}, is a solution of the Einstein equations whose source is an anisotropic fluid. Such a fluid is characterized by the energy density $\rho$, radial pressure $p_r$, and tangential pressure $p_t$ given by the expressions
\begin{eqnarray}
 8\pi\, \rho(r)&=&\dfrac{3}{R^{2}}-\dfrac{5r^{2}}{A^{4}}, \label{eq:density}\\
    8\pi p_r(r)&=&-8\pi \rho(r)=-\left(\dfrac{3}{R^{2}}-\dfrac{5r^{2}}{A^{4}}\right), \label{eq:rpressure}\\
   8\pi p_t(r)&=&-\left(\dfrac{3}{R^{2}}-\dfrac{10r^{2}}{A^{4}}\right). \label{eq:tpressure}
\end{eqnarray}
The energy density given by equation \eqref{eq:density} vanishes at $r_v= \sqrt{\frac{3}{5}}\frac{A^2}{R}$. Therefore, we restrict the analysis of the fluid quantities to the region $r\leq r_v$.

Notice that, while Tolman dealt just with perfect fluid models, the present model represents a nonperfect fluid, i.e., an anisotropic fluid.  In fact, the above solution can be viewed also as a particular case of the anisotropic spheres studied in Ref.~\cite{Maharaj2006}.

With the choice \eqref{eq:Tpotential}, the mass function $M(r)$ is given by
\begin{equation}\label{eq:mfunction}
    M(r)\equiv M_{in}(r)=\dfrac{1}{2}\left(\dfrac{r^{3}}{R^{2}}-\dfrac{r^{5}}{A^{4}}\right), 
\end{equation} 
and then the resulting interior solution is the stationary metric \eqref{eq:RST} with $M(r)$ given by \eqref{eq:mfunction}.

\subsection{The exterior solution}

The exterior spacetime is described by the Kerr metric which, in the Boyer-Lindquist coordinates, is given by \eqref{eq:RST} with a constant mass function, i.e., with 
\begin{equation}
    M(r) = M_{Kerr}= m = {\rm constant.} \label{eq:Kerrmass}
\end{equation}
The exterior Kerr metric is to be smoothly matched to the interior metric considered in Sec.~\ref{Sec:InterSol}.

\subsection{The matching conditions}

The goal of this section is finding the smooth junction conditions between the exterior Kerr metric and the interior axisymmetric metric. For this end we employ the Darmois-Israel~\cite{Israel:1966rt}  matching conditions, see also Drake and Turolla~\cite{Drake:1997hh} for and interesting example of application of these junction conditions to stationary spacetimes. 

As a first step, we need to identify the boundary surface $S$ that separates the interior region from the exterior region of the spacetime. Such a surface is chosen to be $S:r=r_0=$ constant, which is the simplest choice also suggested in Refs.~\cite{Gurses:1975vu,Herrera1982} and \cite{Drake:1997hh}. Hence, the interior region is defined by values of the radial Boyer-Lindquist coordinate $r$ such that $r< r_0$, while the exterior region, where the Kerr metric applies, is for $r$ in the interval $r_0 < r< \infty$.

The next step is to assume that the transition between the two solutions is smooth, what requires that the first fundamental form on $S$ (the projection of the metric tensor at $r=r_0$) and that the second fundamental form (the extrinsic curvature tensor of $S$) be continuous functions across the boundary. Such conditions avoid singular energy-momentum tensor at the matching surface (for more details see Ref.~\cite{Drake:1997hh}). Thus, with the interior and exterior metrics given by the line element~\eqref{eq:RST}, the smooth matching conditions imply the mass function $M(r)$ must be at least $C^{1}$. As a result of these assumptions, the boundary conditions become
\begin{equation}
     \label{eq:boundary}
     \begin{aligned}
      & M_{in}(r_0)= M_{Kerr}(r_0)=m,\qquad \\ & \left.
      \dfrac{d M_{in}(r)}{dr}\right|_{r_0}= \left.\dfrac{d M_{Kerr}(r)}{dr}\right|_{r_0} =0,
\end{aligned}
\end{equation}
where $M_{in}$ stands for the inner mass function, given by Eq.~\eqref{eq:mfunction}, while $M_{Kerr}$ denotes the exterior mass parameter, related to Eq.~\eqref{eq:Kerrmass}. See Appendix \ref{appendixA} for more details.
The two relations in \eqref{eq:boundary} imply the constraints
\begin{align}
   r_{0}^3&=5m\,R^{2}  ,\qquad 
    A^{4}=\dfrac{5r_{0}^{2}R^{2}}{3}, \label{eq:mconstr} 
\end{align}
so that the complete solution  
is given by the line element~\eqref{eq:RST} with the mass function in the form 
\begin{equation}\label{eq:massarot}
 M(r)=\dfrac{5m}{2}\dfrac{r^{3}}{r_0^3}\left(1-\dfrac{3\,r^{2}}{5r_0^{2}}\right)\Theta(r_0-r)+m\,\Theta(r-r_0),
\end{equation}
where $\Theta(x)$ is the step (Heaviside) function.

Notice that the radius of the matching surface coincides with the radius where the energy density  \eqref{eq:density} vanishes. As seen in the previous section, the energy density reaches zero value at $r_v =  \sqrt{\frac{3}{5}}\frac{A^2}{R}$. Moreover, by using the second relation in Eq.~\eqref{eq:mconstr}, it follows $r_v^2 = \frac{3}{5}\frac{A^4}{R^2}= r_0^2$. The vanishing of the energy density exactly at the boundary surface is a consequence of the equation of state, $\rho(r) = - p_r(r)$, and of the smooth boundary conditions imposed here, what implies in $p_r(r_0)=0=\rho(r_0)$. This can be seen from Eqs.~\eqref{eq:density} and \eqref{eq:rpressure} (See also the next section for the analysis of the energy density and pressure of the rotating fluid).

\section{Rotating anisotropic spheroids and rotating regular black holes}\label{Sec:RotAniSph}

\subsection{General properties}

The present model has five constant parameters, namely, $R$, $A$, $r_0$, $m$, and $a$, and two constraints between them given by Eq.~\eqref{eq:mconstr}. Hence, only three of them are free parameters. 
Here we choose to work with $m,\,a,$ and $r_0$ as the relevant free parameters. Moreover, without loss of generality, all parameters carrying dimensions of length may be normalized by $r_0$. Therefore, in the numerical analysis, we take $ m/r_0$, $a/r_0$ as the true free parameters of the model.  

In order to investigate the properties of the solution, we study the behavior of the curvature scalars related to the spacetime metric~\eqref{eq:RST}, with the mass $M(r)$ defined in Eq.~\eqref{eq:massarot}. Besides,  looking for the existence of rotating regular black holes, we investigate the presence of horizons and ergospheres as well. The energy conditions are also analyzed next.

\subsection{Energy-momentum tensor}

Here we follow Ref.~\cite{Gurses:1975vu} to verify that the matter source for the spacetime geometry~\eqref{eq:RST} is an anisotropic fluid that can be cast into the form
\begin{equation}
    T^{\mu\nu}=\epsilon u^{\mu}u^{\nu}+p_{r}e_{r}^{\mu}e_{r}^{\nu}+p_{\theta}e_{\theta}^{\mu}e_{\theta}^{\nu}+p_{\phi}e_{\phi}^{\mu}e_{\phi}^{\nu},
\end{equation}
where $\{u^{\mu},\,e_{r}^{\mu},\,e_{\theta}^{\mu},\,e_{\phi}^{\mu}\}$ is an orthornormal tetrad basis given by 
\begin{equation}
    \begin{split}
        u^{\mu}&=\dfrac{1}{\sqrt{\pm\Delta\Sigma}}\left[(r^{2}+a^{2})\delta_{t}^{\mu}+a\,\delta_{\phi}^{\mu}\right], \\
     e_{r}^{\mu}& =\sqrt{\dfrac{\pm\Delta}{\Sigma}}\delta_{r}^{\mu},\qquad   e_{\theta}^{\mu}=\dfrac{1}{\sqrt{\Sigma}}\delta_{\theta}^{\mu},\\  e_{\phi}^{\mu}&=\dfrac{1}{\sqrt{\Sigma}\,{\rm sin}\,\theta}\left[a\,{\rm sin}^{2}\theta\,\delta_{t}^{\mu}+\delta_{\phi}^{\mu}\right],
    \end{split}
\end{equation}
in which the plus ($+$) sign goes for positive $\Delta(r)$, while the minus ($-$) sign goes for negative $\Delta(r)$, respectively.  In spacetime regions where $\Delta(r)$ is positive, the plus sign is chosen and it results that $u^{\mu}$ is a timelike vector while the remaining are spacelike vectors. In spacetime regions where $\Delta(r)$ is negative, the minus sign has to be chosen and the vectors $u^{\mu}$ and $ e_{r}^{\mu}$ change places, with $ e_{r}^{\mu}$ being the timelike vector of the tetrad.    

The quantities $\epsilon$, $p_r$, $p_{\theta}$, and $p_{\phi}$ are the energy density, the radial, and the tangential pressures of the fluid, respectively, given by the relations 
\begin{equation}\label{eq:RotEnDen}
        8\pi\epsilon=- 8\pi p_r =   
        \dfrac{15m} {r_0^{3}}\dfrac{ r^4}{\Sigma^{2}} \left(1-\dfrac{r^{2}}{r_0^{2}}\right)\Theta(r_0-r),
\end{equation}
\begin{equation}\label{eq:RotPreThe1}
    \begin{split}
        8\pi p_\theta=8\pi p_\phi = &
        \, \dfrac{15m}{r_0^3} \dfrac{r^2}{\Sigma}\left[\dfrac{r^{2}}{\Sigma}\left(1-\dfrac{r^{2}}{r_0^{2}}\right)\right.\\
        & \left. -\left(2-\dfrac{3r^{2}}{r_0^{2}} \right)\right]\Theta(r_0-r).
    \end{split}
\end{equation}

It is straightforward to verify that the energy density $\epsilon$ assumes only nonegative values. For instance, in the equatorial plane ($\theta=\pi/2$) it assumes the maximum value $8\pi\,\epsilon=15m/r_0^3=3/R^{2}$ at $r=0$, and decreases monotonically as $r$ grows, finally vanishing at the surface $r=r_0$. For $\theta\neq \pi/2 $, $\epsilon$ vanishes at $r=0$ and at the surface $r=r_0$, being finite and positive everywhere else in the interval $0< r< r_0$.  
The vanishing of the energy density at the center in the Boyer-Lindquist coordinates is a feature shared by other models of regular black holes found in the literature (see e.g. Refs.~\cite{Toshmatov:2017zpr} and \cite{Smailagic2010} for other examples of this behavior). Alternative mass functions that lead to a nonvanishing central energy density may cause other problems. This is the case of some of the models discussed in Refs. \cite{Bambi:2013ufa, Neves:2014aba}. Another example is the rotating black hole model presented in Ref.~\cite{Toshmatov:2015npp}. It has a central core of quintessential matter with nonzero density at $r\to 0$, but the fluid quantities and the curvature scalars present singularities at the region $r=0,\,\theta\neq \pi/2$.
As shown in Ref.~\cite{Maeda:2021jdc}, the Gürses-Gürsey gives rise to singular spacetimes whenever the mass function is finite or goes to zero at $r\to 0$ as $r^{3+n}$ with $n<1$, which is the case of Ref.~\cite{Toshmatov:2015npp}.

Other interesting conditions that a fluid is usually tested for are the standard energy conditions. Here we need to check just the weak energy condition (WEC).
The WEC applied to a nonisotropic fluid implies in  the constraints $\epsilon\geq 0$ and $\epsilon+p_i\geq0$ ($i=r,\,\theta,\,\phi$) \cite{Hawking:1973uf}. According to Eq.~\eqref{eq:RotEnDen}, the energy density satisfies the first constraint in the whole interval $0\leq r\leq r_0$ for all $\theta$. The second constraint also holds for the radial $p_i=p_r$, i.e., one has $\epsilon+p_r\geq 0$ in the whole interval $0\leq r\leq r_0$ for all $\theta$. On the other hand, as it is seen from Eq.~\eqref{eq:RotPreThe1}, such an inequality does not hold for the tangential pressures $p_i=p_\theta$ and $p_i=p_\phi$. In fact, for $\theta\neq\pi/2$, the constraint $\epsilon+p_\theta\geq 0$ is satisfied in the interval $r_b\leq r\leq r_0$, but it is violated close to $r=0$, in the interval $0 < r < r_b$, where $r_b$ is such that 
$r_b^2 \equiv a^2\cos^2\theta\left( \sqrt{1+ 8r_0^{2}/a^{2}\,\cos^{2}\theta}-1\right)/2$. 

As just analyzed, the present model of rotating compact objects violates the WEC for all $a \neq 0$. This result agrees with Refs.~\cite{Neves:2014aba,Burinskii:2001bq,Torres2017}, where the authors claim that the addition of rotation into regular blackhole solutions inevitably leads to violation of the WEC. This happens for several rotating regular black holes reported in the literature, see Refs.~\cite{Bambi:2013ufa,Toshmatov:2014nya,Dymnikova:2015hka,GHOSH2020115088}. In fact, in Ref.~\cite{Neves:2014aba} it is shown that the black hole rotation unavoidably
leads to the violation of the WEC for any physically reasonable choice of the mass function.
The energy conditions in the Gürses-Gürsey metric has been considered in some detail in the recent work of Ref. \cite{Maeda:2021jdc}\footnote{This paper by K. Maeda appeared in the arXiv just after the publication of first version of the present manuscript in there.}. See also Ref. \cite{Torres:2022twv} for a recent short review on rotating regular black holes.

\subsection{Curvature scalars}

The curvature scalars of the Gürses-Gürsey metric have been analyzed in several works in the literature. Here we mention the recent work by Ref.~\cite{Maeda:2021jdc} where it is shown that, among other interesting results,
when the mass function $M(r)$ is such that it can be expanded as powers of $r$ close to the region $r=0$, i.e.,  $M(r) \simeq r^{3+n}$ with $n\geq 0$ (which is the case here), then curvature singularities are absent. The same result was shown in other previous works, see e.g.  \cite{Neves:2014aba,Torres2017}. 
To verify this result in the present case, we analyze the relevant curvature invariants
(scalars), such as the Ricci scalar $\mathcal{R}=g^{\mu\nu}R_{\mu\nu}$, the Ricci squared $\mathcal{R}_{2} =R_{\mu\nu}R^{\mu\nu}$, and the Kretschmann scalar $\mathcal{K}=R_{\mu\nu\alpha\beta}R^{\mu\nu\alpha\beta}$. For the metric~\eqref{eq:RST},  these curvature scalars take, respectively, the forms  (see also Ref.~\cite{Torres2017} for a complete set of curvature scalars for Petrov type D spacetimes).

\begin{equation}\label{eq:EscCurv}
    \begin{split}
       \mathcal{R}&=\dfrac{2\left(2\,M^{\prime}+r\,M^{\prime\prime}\right)}{\Sigma},\\
       \mathcal{R}_{2}&=\dfrac{1}{\Sigma^{4}}\left[\left(8\,r^{4}+3a^{4}\,{\rm cos}^{4}\,\theta\right)M^{\prime2} \right.\\
       &\left.\quad +r^{4}\,\Sigma^{2}\,M^{\prime\prime2} +4\,r\,\Sigma\,M^{\prime}\,M^{\prime\prime}\,a^{2}\,{\rm cos}^{2}\,\theta\right],\\
       \mathcal{K}&=\dfrac{4}{\Sigma^{6}}\left[384\,r^{6}\,M^{2}-192\,r^{4}\,M\,\Sigma\left(3\,M+r\,M^{\prime}\right) \right.\\
       &\quad+\Sigma^{4}\left(2\,M^{\prime}+r\,M^{\prime\prime}\right)^{2}+8\,r^{2}\,\Sigma^{2}\left(27\,M^{2}\right.\\
       &\left.\quad+4\,r^{2}\,M^{\prime2}+28\,r\,M\,M^{\prime}+2\,r^{2}\,M\,M^{\prime\prime}\right)\\
       &\quad-4\,\Sigma^{3}\left(12\,r\,M\,M^{\prime}+3\,M^{2}+3\,r^{2}\,M\,M^{\prime\prime}\right.\\
       &\quad\left.+7\,r^{2}\,M^{\prime2}+2\,r^{3}\,M^{\prime}\,M^{\prime\prime}\right)\big].
    \end{split}
\end{equation}

The Kerr solution is recovered by taking $M'(r)=0$, which implies that ${\cal R}$ and ${\cal R}_2$ are identically zero, but the Kretschmann scalar is not. In the case of the Kerr spacetime, this scalar diverges at the location where the function $\Sigma=r^{2}+a^{2}\,{\rm cos}^{2}\,\theta$ vanishes, i.e., at $(r,\,\theta)=(0,\,\pi/2)$, what represents a ring-shaped singularity. In other situations, curvature singularities may also occur at the spacetime region where $\Sigma(r,\theta)$ vanishes as it happens for the Kerr metric, even when $M(r)$ and its derivatives are required to be regular functions. However, a careful analysis shows that this is not the case for the metric \eqref{eq:RST} with the mass function given by Eq.~\eqref{eq:massarot}. In fact, by approaching the region $r=0$  along the equatorial plane, i.e., by putting $\theta =\pi/2$ and then taking the limit $r\to 0$, the scalar invariants \eqref{eq:EscCurv} reduce, respectively, to  
\begin{equation}\label{eq:CurInvEP}
    \begin{split}
        \lim_{r\to0} &\left(\lim_{\theta\to\pi/2}\,\mathcal{R}\right)=\dfrac{12}{R^{2}}=\dfrac{60\, m}{r_0^3},\\
        \lim_{r\to0} &\left(\lim_{\theta\to\pi/2}\,\mathcal{R}_{2}\right)=\dfrac{36}{R^{4}}= \dfrac{900\, m^2}{r_0^6},\\
        \lim_{r\to0} &\left(\lim_{\theta\to\pi/2}\,\mathcal{K}\right)=\dfrac{24}{R^{4}}=\dfrac{600\, m^2}{r_0^6}.
    \end{split}
\end{equation}
Additionally, by approaching the region $r=0$ from outside the equatorial plane, that is, $r\to 0$ with $\theta \neq\pi/2$, we have that all the curvature invariants vanish. In fact, for small $r$ we have $M(r)= {\cal O}\left(r^3\right)$, $ M'(r)= {\cal O}\left(r^2\right)$, and  $M''(r)={\cal O}\left(r\right)$. Then it follows
\begin{equation}\label{eq:CurInvEP2}
    \begin{split}
        \lim_{\theta\to \pi/2} &\left(\lim_{\to0}\,\mathcal{R}\right)=\lim_{\theta\to \pi/2}  \left(\lim_{r\to0}\frac{{\cal O}\left(r^2\right)}{\Sigma}\right)=0,\\
        \lim_{\theta\to \pi/2} &\left(\lim_{r\to 0}\,\mathcal{R}_{2}\right)=\lim_{\theta\to \pi/2}  \left(\lim_{r\to0}\frac{{\cal O}\left(r^4\right)}{\Sigma^3}\right)=0,\\
        \lim_{\theta\to \pi/2} &\left(\lim_{r\to 0}\,\mathcal{K}\right)=\lim_{\theta\to \pi/2}  \left(\lim_{r\to0}\frac{{\cal O}\left(r^6\right)}{\Sigma^6}\right)=0.
    \end{split}
\end{equation}
This means that the curvature invariants have finite values everywhere in the Boyer-Lindquist coordinates, but they are not defined on the ring $(r = 0,\,\theta = \pi/2 )$.  Therefore, we conclude that, differently from the Kerr metric, the spinning geometry generated by the mass function \eqref{eq:massarot} is free of curvature singularities.

\subsection{Horizons}

Even though the presence of horizons in the Gürses-Gürsey metric for several mass functions  $M(r)$ has been considered in previous studies, it is worth presenting here some aspects of the horizon function for the particular case of $M(r)$ given by Eq.~\eqref{eq:massarot}. Such aspects are useful for the analysis of the causal structure presented below. 

\begin{figure*}[ht]
         \includegraphics[width=0.32\textwidth]{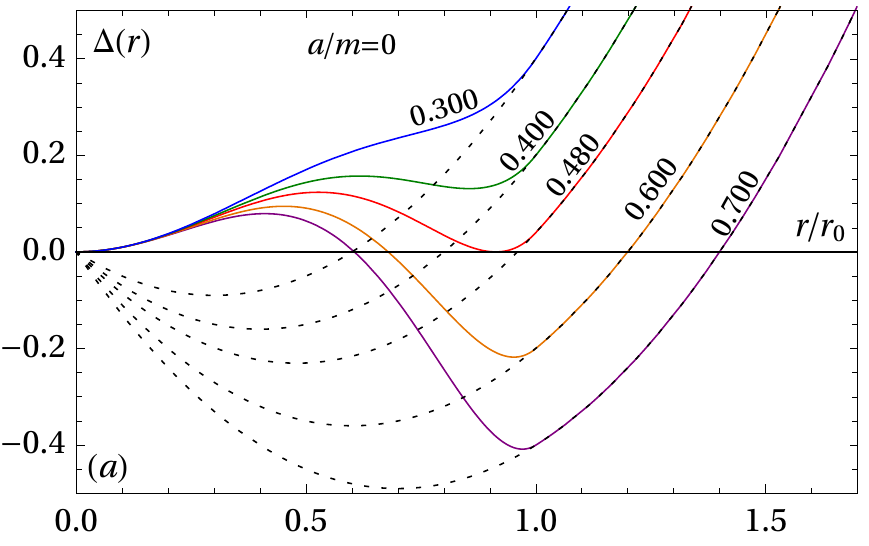}
         \includegraphics[width=0.32\textwidth]{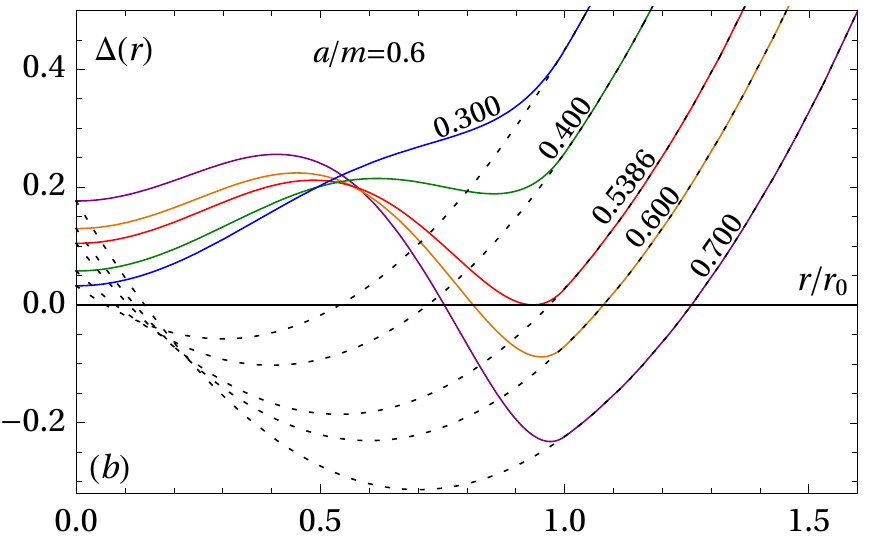}
         \includegraphics[width=0.31\textwidth]{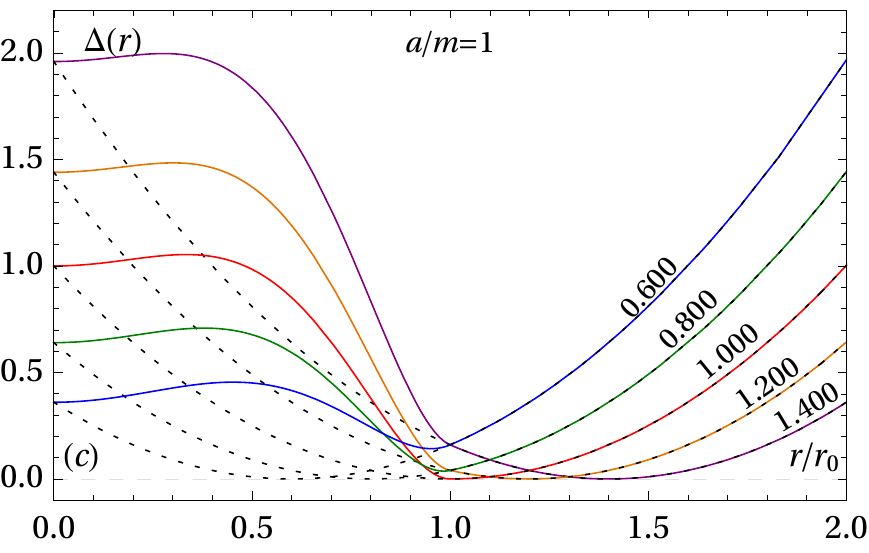}
        \caption{The horizon function $\Delta(r)$ as a function of $r/r_0$ for different values of the total mass $m/r_0$ (as indicated by the label on each curve), and for three values of the rotation parameter $a/m$: $a/m=0$, $a/m=0.6$ and $a/m=1$. Each solid line represents a complete solution, while the dotted line represents the corresponding Kerr solution. The zeros in $\Delta(r)$ indicate the existence of horizons. }
        \label{fig:Deltarxa}
\end{figure*}

In Boyer-Lindquist coordinates, metric~\eqref{eq:RST} presents horizons at the spacetime regions where the function $\Delta(r)$ vanishes. Thus, from Eq.~\eqref{eq:Delta} it follows that the horizons radii $r_i$ are the real and positive roots of the polynomial equations
\begin{equation} \label{eq:horizons}
\begin{split}
    & a^2 +r^2 -\dfrac{5m}{r_0}\left(1-\dfrac{3\,r^{2}}{5r_0^{2}}\right)\dfrac{r^{4}}{r_0^2} =0,\quad r<r_0,\\ 
    &  a^2 +r^2 -2m\, r =0,\qquad r\geq r_0,
\end{split}
\end{equation} 
with the supplemental conditions that the horizons are determined by the solutions of the first equation when $r_i\leq r_0$, but are given by the solutions of the second equation when $r_i\geq r_0$.  

A simple analysis shows that each one of the polynomial equations \eqref{eq:horizons} may have two, one, or none real positive roots, depending on the relative values of the free parameters $a$, $m$, and $r_0$.  
The critical case that separates situations with no horizon from situations with two horizons is the extremal case, in which $r_-=r_+\equiv r_c$. In such a case, Eq.~\eqref{eq:horizons} yields the extremal mass $m_c$, 
\begin{eqnarray} 
    \dfrac{m_c}{r_0}&\! =\!\!  & \dfrac{1}{1000}\dfrac{r_0^2}{a^2} \sqrt{\left(5 + \dfrac{3a^2}{r_0^2}\right)\left(5 +\dfrac{27a^2}{r_0^2}\right)^{\!\!3}}\nonumber \\ & &+ \dfrac{243}{1000}\dfrac{a^2}{r_0^2} -\dfrac{25}{1000}\dfrac{r_0^2}{a^2}+ \dfrac{27}{100}, \quad \label{eq:CritMass}
    \end{eqnarray}
for $ \dfrac{r_c}{r_0} < 1$, and  
\begin{eqnarray}
    \dfrac{m_c^2}{r_0^2} &= & \dfrac{a^2}{r_0^2}, \qquad \mathrm{ for}\quad \dfrac{r_c}{r_0} >1. \label{eq:critmass2}
\end{eqnarray}

The extremal mass given by the first solution, i.e., given by \eqref{eq:CritMass}, applies to the cases in which the double horizon is located in the inner region of the spacetime. 
This first solution presents a minimum extremal mass given by $m_c/r_0=12/25=0.48$ at $a=0$. This means that, differently from the Kerr metric, rotating extremal (double horizon) black holes may occur even for rotation parameter smaller than unity. 
 The second solution, given in \eqref{eq:critmass2}, applies to the cases in which the double horizon is located in the exterior region. 
   More details on this subject are reported  in Sec.~\ref{sec:characterS}. 

Figure~\ref{fig:Deltarxa} shows the behavior of $\Delta(r)$ as a function of $r/r_0$ for three different values of the ratio $a/m$, namely, $a/m=0$, $a/m=0.6$, and $a/m=1.0$, and five different values of the mass to radius ratio $m/r_0$. The dotted lines are associated to the exterior Kerr metric, and are drawn for constant $M(r)$, i.e., for $M(r)=m$, while the solid lines correspond to the complete solution with $M(r)$ given by \eqref{eq:massarot}.
As seen from the figures, the solid and dotted lines are the same in the vacuum region, i.e., for $r/r_0\geq 1$.  The intersections of each solid curve with the horizontal axis determine the existence and the position of the horizons $r_-$, the smallest root that corresponds to the inner (Cauchy) horizon, and $r_+$, the largest root that corresponds to the event horizon. 

\begin{figure*}[hbt] 
         \includegraphics[width=0.32\textwidth]{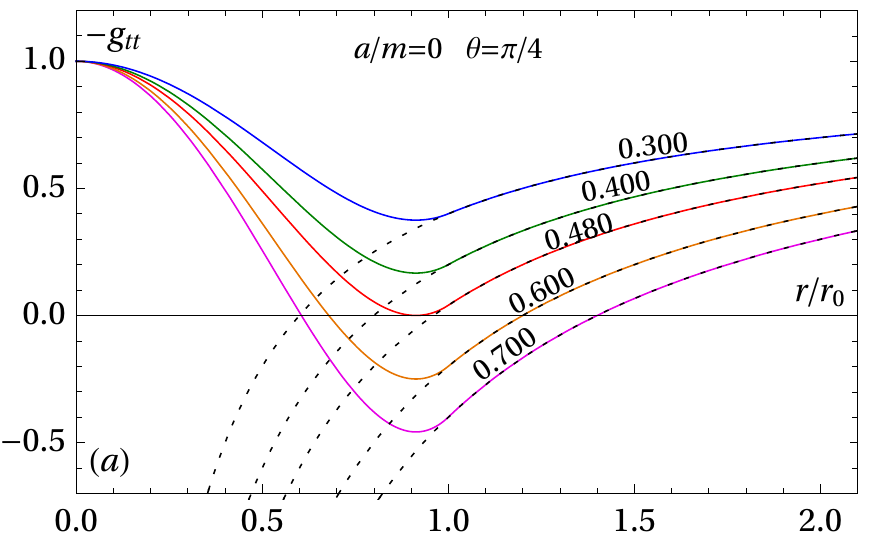}
       \includegraphics[width=0.32\textwidth]{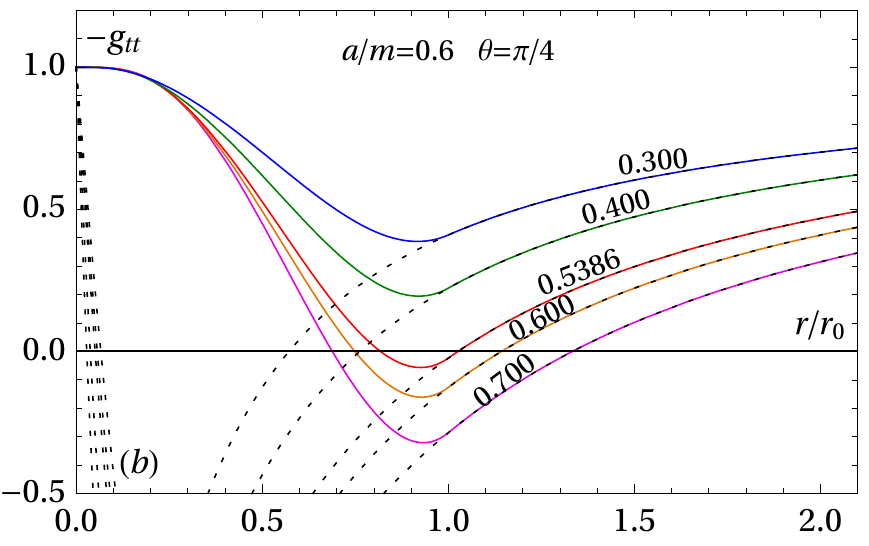}
         \includegraphics[width=0.32\textwidth]{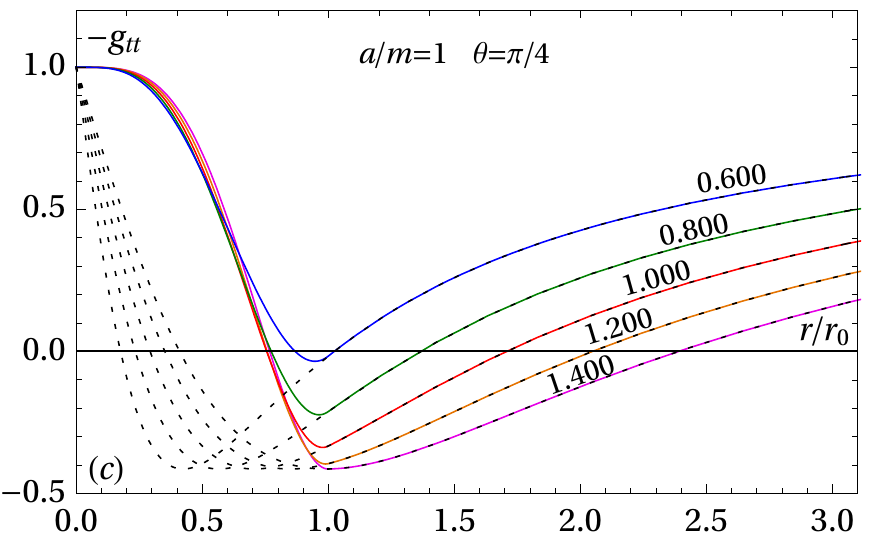}
        \caption{The behavior of the metric coefficient $g_{tt}$ as a function of $r/r_0$ on the transverse plane $\theta=\pi/4$, for different values of the total mass $m/r_0$ (as indicated by the label on each curve), and for three values of the rotational parameter $a/m$, namely, $a/m=0$, $a/m=0.6$, and $a/m=1$. The solid lines represent the complete solutions, while the dotted lines represent the corresponding Kerr solution. The zeros of $g_{tt}$ indicate the surfaces of static limit (the ergosurfaces).}
        \label{fig:gtt}
\end{figure*}
The effects of rotation on the horizon radii may be seen by comparing panel (a) to the other two panels of the same figure. 
An interesting effect is the increase in the maximum masses of solutions representing regular stars.

\subsection{Ergosurfaces and ergoregions}

Other important surfaces when considering stationary spacetimes are the surfaces of static limit, the so-called ergosurfaces. 
In the present case, such limiting surfaces correspond to spacetime regions where the Killing vector $\chi^{\mu}=(1,\,0,\,0,\,0)$ becomes lightlike. This condition implies in $g_{tt}=0$, leading to the following equations,
\begin{equation}\label{eq:ergosurfs}
\begin{split}
  & a^{2}\cos^{2}\theta +r^2  -\dfrac{5m}{r_0}\left(1-\dfrac{3\,r^{2}}{5r_0^{2}}\right)\dfrac{r^{4}}{r_0^2} =0, \quad r< r_0,\\
   & a^{2}\cos^{2}\theta +r^2  -2m\, r  =0, \qquad r\geq r_0.
     \end{split}
\end{equation} 
As in the case of the Kerr geometry, these equations may have two real positive roots that define two ergosurfaces that delimit the ergoregions. We denote the two roots by $r_{e^\pm}$. 
 
The dependency of the metric coefficient $g_{tt}$ defined in Sec.~\ref{Sec:RotTol} as a function of the normalized radial coordinate $r/r_0$ is shown in Fig.~\ref{fig:gtt}. Each panel is drawn for $\theta = \pi/4$, for a different value of the relative rotation parameter $a/m$, and for a few values of the total mass $m/r_0$, as indicated by the line labels. The conventions are the same as in Fig.~\ref{fig:Deltarxa}. 

The inclusion of rotation makes the surfaces of stationary limit to differ from the horizons, as verified by comparing the curves in  panels (a), (b), and (c) of Fig.~\ref{fig:gtt} with the corresponding curves in panels (a), (b), and (c) of Fig.~\ref{fig:Deltarxa}, respectively. Also, it is easy to verify that the ergosurfaces may be present in cases where horizons are absent. This happens for rotating configurations with sufficiently high values of the rotation parameter $a/m$ and for masses close to the extremal mass $m_c$.

 The ergosurfaces of the present stationary spacetime solutions are also displayed in Figs.~\ref{fig:Ergo0}, \ref{fig:Ergo}, and \ref{fig:Ergo1} by solid blue lines for some values of the parameter $a/m$. The matching surfaces and the horizons (when present) are also shown in that figures respectively by dashed black lines and dotted red lines. A more detailed description of these figures is presented next.

\subsection{On the causal character of the matching surface and classification of the solutions}
\label{sec:characterS}

\subsubsection{On the causal character of the matching surface}
\label{sec:characterS1}

After having determined the horizons and ergoregions, the study of some properties of the matching surface gets simplified, and we turn attention to this subject here. The junction between the two spacetime regions may be of three different kinds, according to the causal character of the  matching surface being timelike, lightlike or spacellike. Such a character may be determined by calculating the normal vector to the boundary surface $n^\nu$, whose norm is $N= \Sigma\, \Delta/\Upsilon$ (see Appendix \ref{appendixA}).  Hence, in order to determinate the kind of matching, one needs to find the particular region in the parameter space where the norm $N$ changes of sign. Since the functions $\Sigma$ and $\Upsilon$ are non-negative in the region of interest, the causal character of the matching surface is fully determined by the function $\Delta(r)$ alone.
The critical values of the parameters are calculated by the condition $\Delta(r=r_0)=0$, what furnishes a critical mass given by
\begin{equation}\label{eq:masshor}
    \dfrac{m_{h}}{r_0}=\dfrac{1}{2}\left(1+ \dfrac{a^{2}}{r_0^{2}}\right), 
\end{equation}
which represents a line in the parameter space spanned by $m/r_0$ and $a/r_0$. Therefore, 
 objects whose masses $m$ equal the critical mass $m_h$ have a lightlike boundary surface.
Additionally, the function $\Delta$ is positive for objects with smaller masses, $m <m_h$, in which cases the matching surface is timelike, while it is negative for larger masses, $m>m_h$, in which cases the matching surface is spacelike.

 Note that extremal masses depend just on the ratio $a^2/r_0^2$, and that the inequality $m_c/m_h\leq 1$ holds for all values of $a/r_0$, with the equality being valid at $a^2/r_0^2=1$. In fact, the first solution for $m_c$ given by Eq.~\eqref{eq:CritMass}, that holds for $r_0<r_c$, is such that the ratio $m_c/m_h$ is restricted to the interval $0. 96\leq m_c/m_h\leq 1$  for $a^2/r_0^2\in [0,\, 1]$, and is restricted to $1\geq  m_c/m_h\geq 0.972$ for $a^2/r_0^2\in [1,\, \infty)$.
In turn, the second solution for $m_c$ given by \eqref{eq:critmass2}, that holds for $r_0>r_c$, is such that the ratio $m_c/m_h$ is restricted to the interval $0\leq m_c/m_h\leq 1$ for $a^2/r_0^2\in [0,\, 1]$, and is restricted to $1\geq m_c/m_h\geq 0$ for $a^2/r_0^2\in [1,\, \infty)$.

The properties of the matching surface in terms of the free parameters of the model $a/r_0$ and $m/r_0$ are considered next. The analysis is split into three distinct cases depending on the values of the normalized rotation parameter $a/m$, namely, $0\leq a/m<1$, $a/m=1$, and $a/m>1$.

\begin{figure*}[htb] 
        \includegraphics[width=0.24\textwidth]{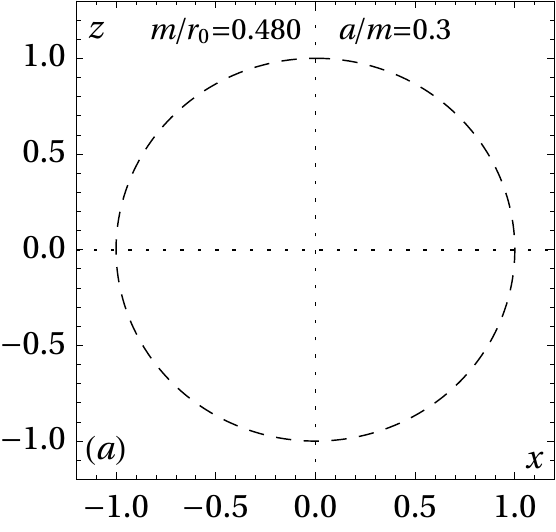}
        \includegraphics[width=0.24\textwidth]{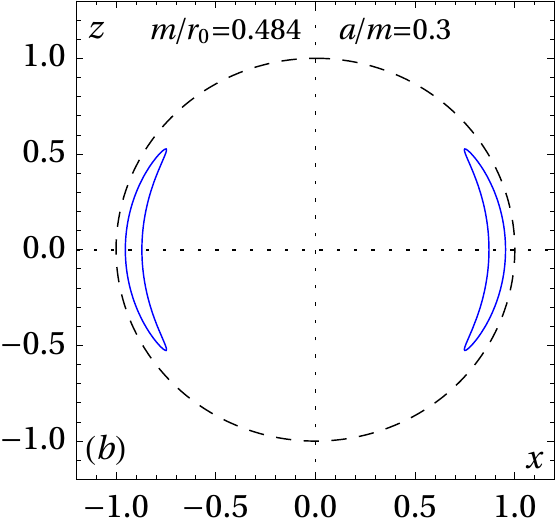}
        \includegraphics[width=0.24\textwidth]{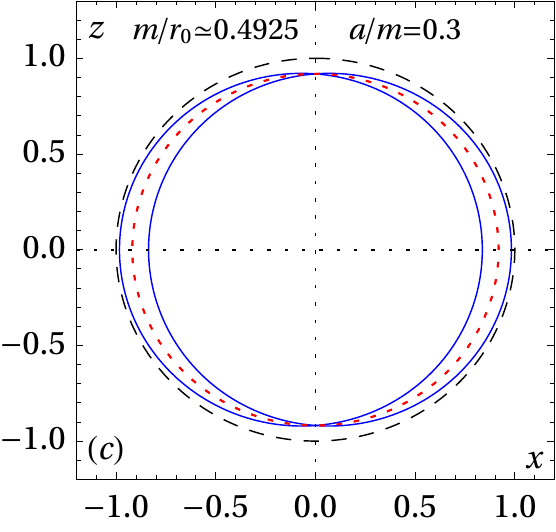}
         \includegraphics[width=0.24\textwidth]{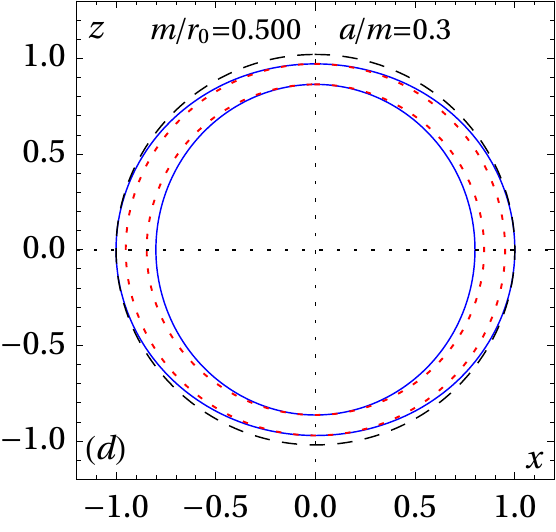}
        \caption{Ergospheres, horizons, and matching surfaces in the $x$--$z$ plane, where $x=(r/r_0)\sin{\theta}$ and $z=(r/r_0)\cos{\theta}$, for $a/m=0.3$ and three values of $m/r_0$, as indicated by the labels in each plot. The solid blue lines represent the ergosurfaces, the dotted red lines are the horizons, and the dashed black lines indicate the junction surfaces at $r=r_0$.}
         \label{fig:Ergo0}
\end{figure*}

\begin{figure*}[htb] 
        \includegraphics[width=0.245\textwidth]{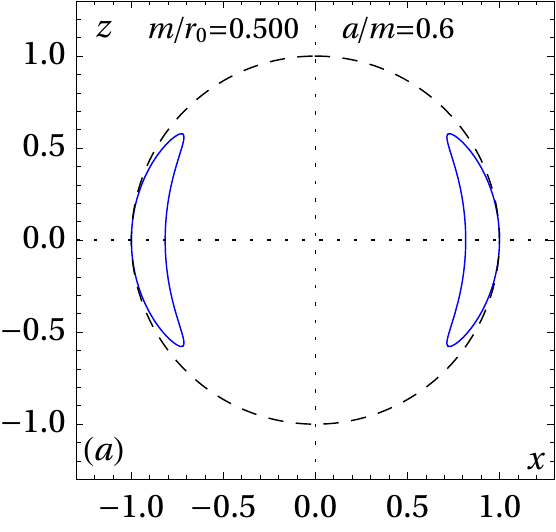}
        \includegraphics[width=0.245\textwidth]{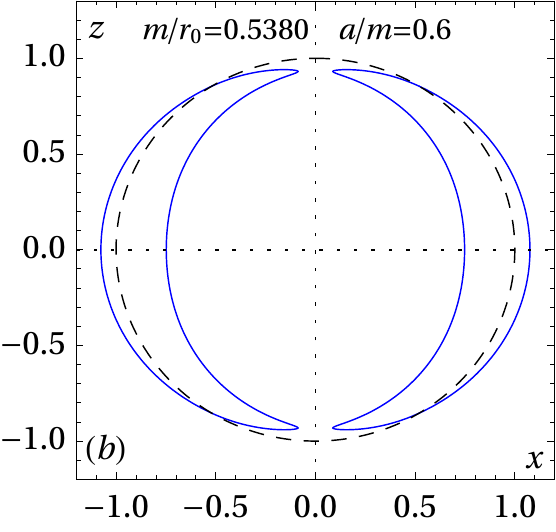}
         \includegraphics[width=0.245\textwidth]{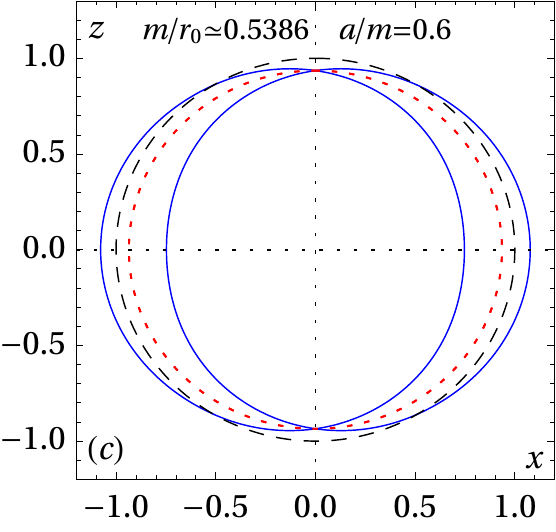}
%
         \includegraphics[width=0.245\textwidth]{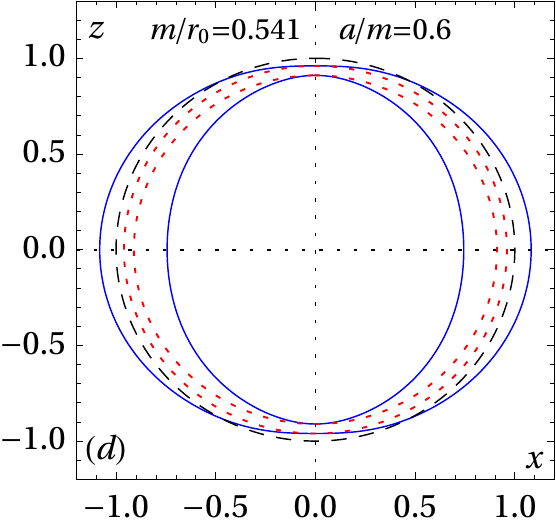}
         \includegraphics[width=0.247\textwidth]{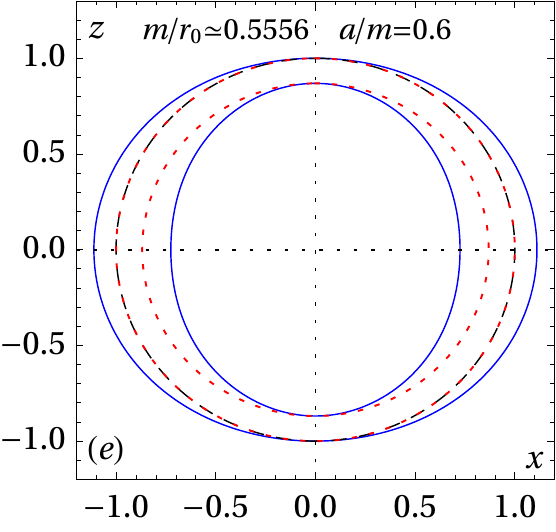}
         \includegraphics[width=0.247\textwidth]{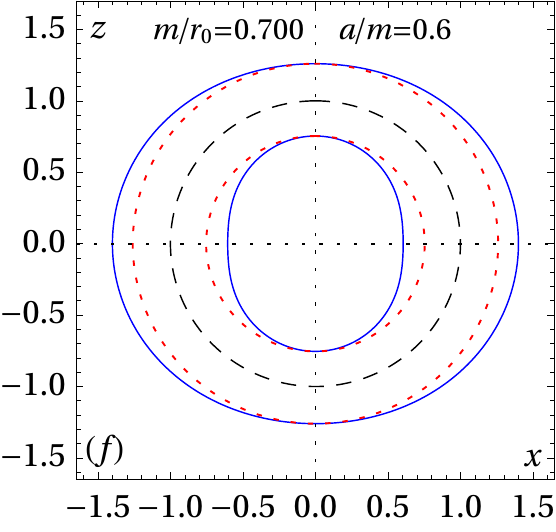}
         \includegraphics[width=0.243\textwidth]{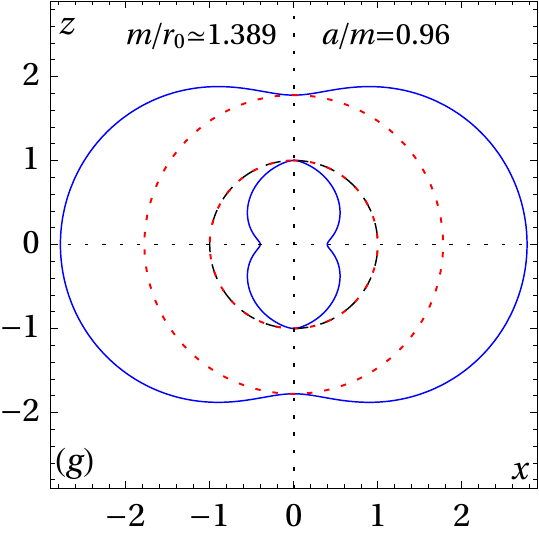}
         \includegraphics[width=0.243\textwidth]{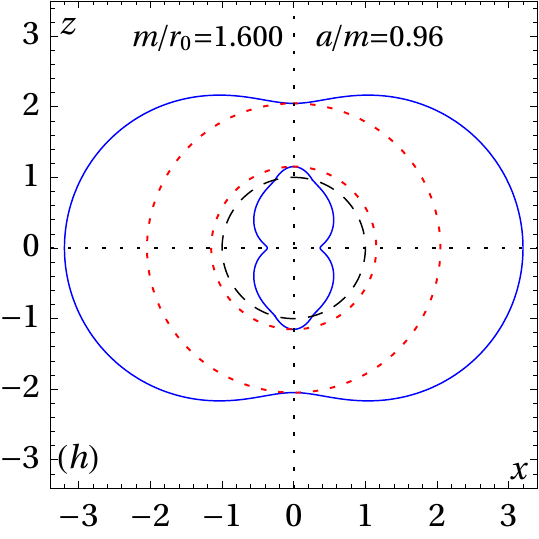}
        \caption{Ergospheres, horizons, and matching surface in the $x$--$z$ plane, where $x=(r/r_0)\sin{\theta}$ and $z=(r/r_0)\cos{\theta}$. The solid blue lines represent the ergosurfaces, the dotted red lines are the horizons, and the dashed black lines indicate the junction surfaces at $r=r_0$. The first six plots are drawn for $a/m=0.6$. The last two plots are drawn for $a/m=0.96$, i.e., with a larger rotation parameter to better visualize the cases where the matching surface is inside the Cauchy horizon or coincides with it.}
        \label{fig:Ergo}
\end{figure*}

\subsubsection{Underspinning objects: For $0\leq a/m<1$}
\label{sec:characterS2}

This is the richest region of the space of parameters because it presents several classes of objects. The different classes depend on the relative values of the masses of the objects in comparison to the extremal mass $m_c$ defined in Eqs. \eqref{eq:CritMass} and \eqref{eq:critmass2}, and to the critical mass $m_h$ defined in Eq.~\eqref{eq:masshor}. In fact, twelve different configurations are seen in Figs.~\ref{fig:Ergo0} and~\ref{fig:Ergo}. 
For the sake of convenience we split the description of these objects into nine cases, as follows. \\

 \noindent
{\it (i) For small masses, $0<m/r_0\leq 12/25=0.48$}:

Independently of the values of the rotation parameter  $a/r_0$, for masses such that $0<m/r_0\leq 12/25=0.48$, the solutions present neither horizons nor ergospheres, and the corresponding configurations are rotating (static in the case $a=0$) regular stars where the matching is done on a timelike surface. Typical examples of these type of configurations are shown by the upper curves (solid blue and green lines) in Figs.~\ref{fig:Deltarxa} and \ref{fig:gtt}, and in panel (a) of Fig.~\ref{fig:Ergo0}. \\

\noindent
{\it (ii) For masses in the interval $12/25<m/r_0<m_c/r_0$}:

Independently of the values of the rotation parameter $a/r_0$,  the solutions present no horizons but ergospheres are formed. The configurations represent regular stars with timelike matching. Within this interval of masses, depending on the value of $m_c/r_0$ being smaller or larger than $1/2$, two distinct configurations of ergospheres are found, as follows.

\vskip .1cm\noindent
{\it (ii.a) For $m_c/r_0\leq 1/2$}: 

This condition also means $12/25<m/r_0<m_c/r_0\leq 1/2$ and implies in the constraint $0<a/r_0<\dfrac{1}{9}\sqrt{\left(7\sqrt{7}-10\right)/3}\simeq0.1873$. In such a case there are two disjoint ergoregions located completely inside the matter distribution, $r_{e\pm}<r_0$. A typical example of this kind of configurations is shown in panel (b) of Fig.~\ref{fig:Ergo0}. In the limiting case, $m/r_0=1/2$, the external ergosurface ($r_{e^+}$) reaches the matching surface $r_0$ at the equatorial plane. A typical example of this kind of configurations is drawn in panel (a) of Fig.~\ref{fig:Ergo}.

\vskip .1cm \noindent
{\it (ii.b) For $m_c/r_0> 1/2$}:

This condition also means $1/2<m/r_0<m_c/r_0$ and implies in the constraint $\dfrac{1}{9}\sqrt{\left(7\sqrt{7}-10\right)/3}0< a/r_0$. In such a case the external ergosurface $r_{e^+}$ reaches the region outside matter close to the equatorial plane. A typical example of this kind of configurations is drawn in panel (b) of Fig.~\ref{fig:Ergo}.\\

\noindent
{\it (iii) The extremal case, for $m/r_0=m_c/r_0$}: 

In this case the two horizons coincide and the matching surface can be timelike or lightlike, depending on the value of $m_c/r_0$ being smaller, equal, or larger than 1. Three distinct configurations are found, as follows.

\vskip .1cm\noindent
{\it (iii.a)  For $m/r_0=m_c/r_0$ with $m_c/r_0<1$}: 

In the extremal case and for configurations with masses such that $m/r_0<1$, which also obeys $a/r_0<1$ and $0 < a/m< 1$, the two horizons coincide, the matching surface is timelike and is located outside the horizon, $r_0>r_+=r_-$. The double horizon and the ergosurfaces coincide at the poles. The two ergosurfaces join each other at the poles forming a unique ergoregion. Close to the poles,  the exterior ergosurface is located inside the matter distribution ($r_{e^+}<r_0$), but it may enter the region outside the matter distribution ($r_{e^+}>r_0$) close to the equator. The corresponding spacetimes are extremal rotating regular black holes. A typical case where the double horizon and the ergoregions are formed inside the matter region is 
shown in panel (c) of Fig.~\ref{fig:Ergo0}.
A typical case of the situation where the double horizon is located inside the matter region, while the outer ergosurface enters the exterior region is shown by the curve labeled with $m/r_0=0.5386$ in panels (b) of Figs.~\ref{fig:Deltarxa} and~\ref{fig:gtt}. The corresponding ergosurfaces and horizons are shown in panel (c) of Fig.~\eqref{fig:Ergo}.

\vskip .1cm \noindent
{\it (iii.b) For $m/r_0=m_c/r_0$ with $m_c/r_0=1$}: 

In the case $m/r_0=m_c/r_0=1$, which also obeys $a/r_0=1$, the matching surface is lightlike and it is located at the double horizon, $r_0=r_-=r_+$. A more detailed description on this case is presented in Sec.~\ref{sec:characterS3}.

\vskip .1cm \noindent
{\it (iii.c) For $m/r_0=m_c/r_0$ with $m_c/r_0 > 1$}:

The extremal cases with $m/r_0=m_c/r_0\geq 1$ also obey $m/r_0=a/r_0 >1$, the matching surface is timelike and it is located inside the double horizon, $r_0<r_-=r_+$, so that Eq.~\eqref{eq:critmass2} provides the relation between the extremal mass $m_c$ and the rotation parameter $a$. I.e., this kind of configurations occurs for $a/m=1$ and then a more detailed description is presented in Sec.~\ref{sec:characterS3}.\\

\noindent
{\it (iv) For intermediate masses, $m_c/r_0\leq m/r_0 <m_{h}/r_0$}: 

In models with intermediate masses, when the total mass takes values in the range $m_c/r_0\leq m/r_0 <m_{h}/r_0$, two horizons appear and the matching is done on a timelike surface located at a radius $r_0$ larger than the event horizon radius, $r_0>r_+$. The two horizons are placed inside the inner region and the geometries correspond to rotating regular black holes. Typical examples of this kind of configurations are shown in panels (d) of Figs.~\ref{fig:Ergo0} and \ref{fig:Ergo}. For $m_c/r_0 < m/r_0 \leq 0.5$, the ergosurfaces are completely inside the matching surface, cf. panel (d) of Fig.~\ref{fig:Ergo0}.  For $m_c/r_0 < m/r_0 > 0.5$, the outer ergosurfaces reaches the region outside matter,  cf. panel (d) of Fig.~\ref{fig:Ergo}. \\

\noindent
{\it (v) For the critical mass, $m/r_0=m_h/r_0$}:

In this critical case the matching surface is lightlike and it is located at one of the horizons, i.e., at the event horizon $r_0=r_+$, or at the Cauchy horizon $r_0=r_-$. The case $r_0=r_+$ occurs for $m_h/r_0<1$, in which the Cauchy horizon $r_-$ is located in the inner (matter) region of the spacetime. The case $r_0=r_-$ occurs for $m_h/r_0>1$ so the matching surface coincides with the Cauchy horizon. In both cases the solutions represent rotating regular black holes. 
A typical example of configurations with $r_0=r_+$ is shown in panel (e) of Fig.~\ref{fig:Ergo}, while a typical example of configurations with $r_0=r_-$ is shown in panel (g) of Fig.~\ref{fig:Ergo}. \\

\noindent
{\it (vi) For $m/r_0>m_{h}/r_0$ with $m_h/r_0 < 1$}:  

In the case of larger masses, when $m/r_0>m_{h}/r_0$ but with $m_h/r_0<1$, which also obeys $0 < a/r_0<1$, the matching is done on a spacelike surface located between the two horizons, $r_-<r_0<r_+$. The Cauchy horizon is located in the matter region and the solutions correspond to rotating regular black holes. 
A typical case of this kind of configurations is shown by the curve labeled with $m/r_0=0.700$ in panel (b) of Fig.~\ref{fig:Deltarxa}, and in panel (f) of Fig.~\ref{fig:Ergo}. \\

\begin{figure*}[ht] 
         \includegraphics[width=0.25\textwidth]{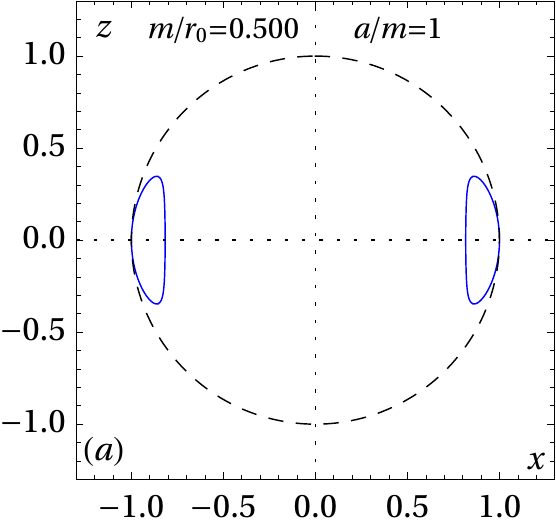}
        \includegraphics[width=0.25\textwidth]{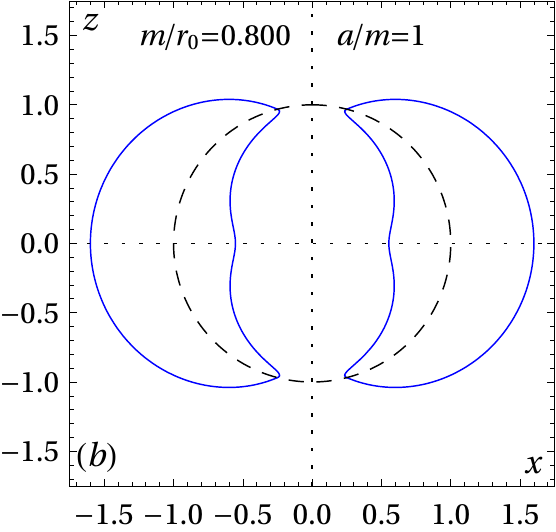}
         \includegraphics[width=0.24\textwidth]{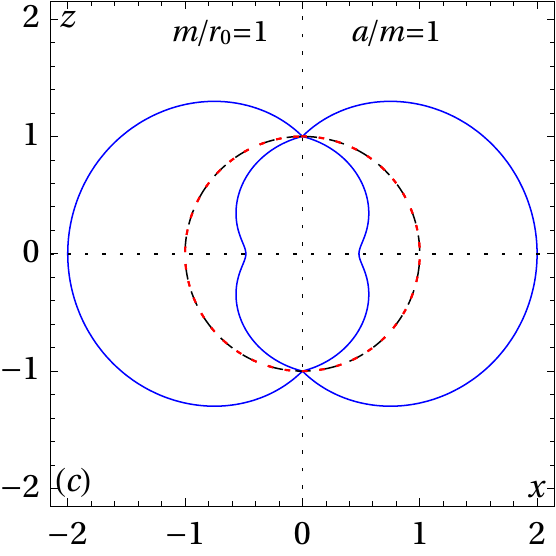}
         \includegraphics[width=0.24\textwidth]{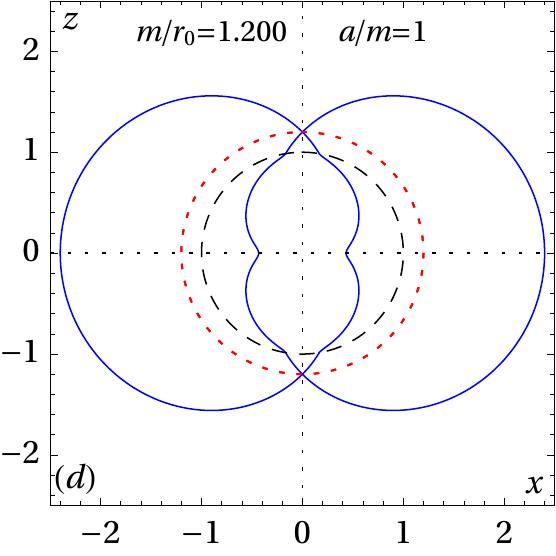} 
                 \includegraphics[width=0.25\textwidth]{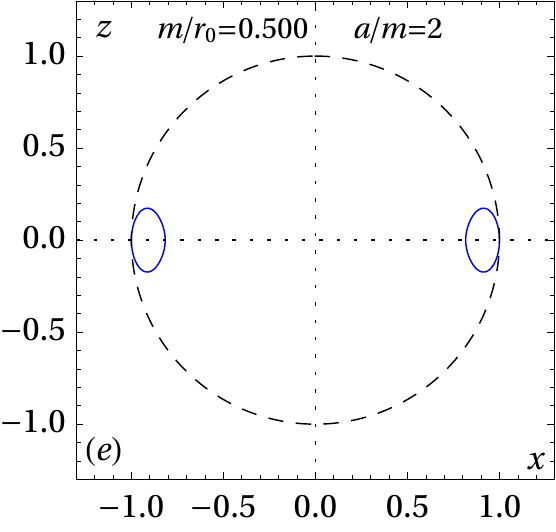}
        \includegraphics[width=0.25\textwidth]{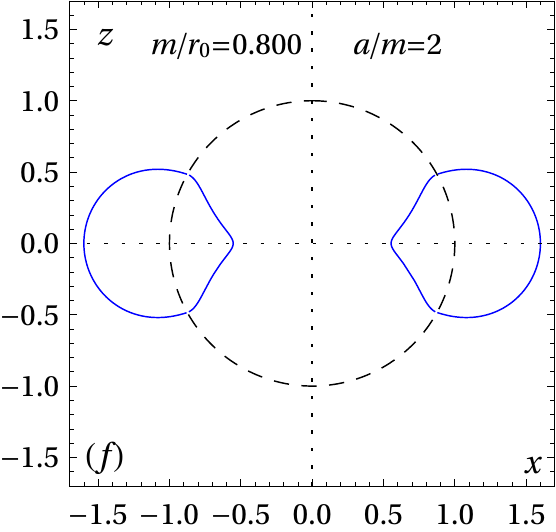}
        \includegraphics[width=0.24\textwidth]{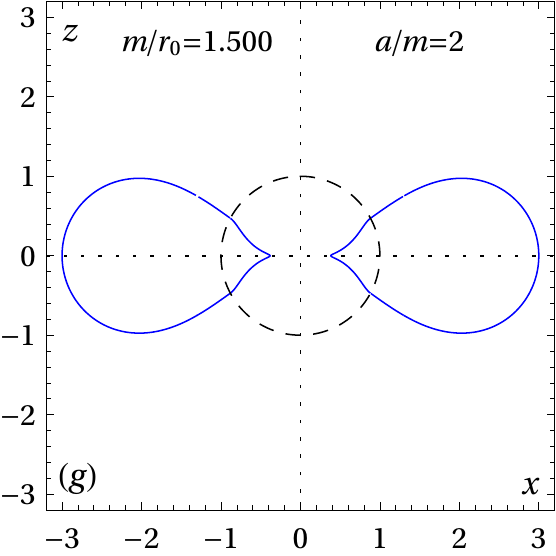}
%
        \includegraphics[width=0.24\textwidth]{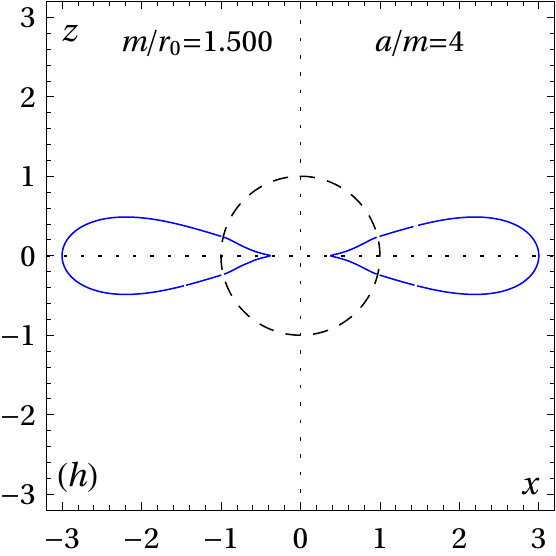}
        \caption{Ergospheres, horizons, and the matching surface for the extremal rotation $a/m=1$ (top panels) and for higher rotation (bottom panels) in the $x$--$z$ plane, where $x=(r/r_0)\sin{\theta}$ and $z=(r/r_0)\cos{\theta}$. The solid blue lines represent the ergosurfaces, the dotted red lines are the horizons, and the dashed black lines indicate the boundary surfaces at $r=r_0$.}
        \label{fig:Ergo1}
\end{figure*}

\noindent
{\it (vii) For $m/r_0>m_{h}/r_0$ and $m_h/r_0> 1$}:

In the case of very large masses, when $m/r_0>m_{h}/r_0$ and with $m_h/r_0> 1$, which also means $a/r_0>1$, the matching is done on a timelike surface located inside the Cauchy horizon, $r_0<r_-<r_+$. The two horizons are placed outside the matter region and the solutions correspond to rotating regular black holes. A typical case of this kind of configurations is shown in panel (h) of Fig.~\ref{fig:Ergo}.\\

\noindent
{\it (viii) Further comments o some special cases}:

Finally we notice that there are a few particular configurations with special characteristics that occur for small values of the rotation parameters, more specifically, for $a/m$ in the interval $0 < a/m \leq \dfrac{2}{9}\sqrt{\left(7\sqrt{7}-10\right)/3}\simeq0.3745$. The reason for the appearance of special cases in this region of the parameter space is related to the smallness of the extremal mass $m_c/r_0$, cf. Eq.~\eqref{eq:CritMass}.  In fact, 
for configurations with rotation parameters in the interval $0<a/m\leq 0.3745$, the minimal extremal masses are small and bounded by $12/25< m_c/r_0\leq 1/2$ so that the horizons and the ergosurfaces may be formed entirely within the matter region. A few representative configurations of this kind are shown in Fig.~\ref{fig:Ergo0}. For masses in the interval $0<m/r_0\leq 12/25$, the solutions are similar to the cases with larger $a/m$ commented in item \textit{(i)} above, in this section, and correspond to rotating stars without ergoregions. For $12/25<m/r_0<m_c/r_0<1/2$, the ergosurfaces are formed entirely inside the matter region, representing regular star models similar to the cases studied in item \textit{(ii)} of the last subsection. In cases with masses constrained by $m_c/r_0\leq m/r_0\leq 1/2$, the horizons and the ergoregions are formed entirely inside the matter region, this is seen in panels (c) and (d) of Fig.~\ref{fig:Ergo0}. The cases with larger masses, i.e., for $m/r_0>1/2$, are similar to the cases shown in panels (d)--(g) of Fig.~\ref{fig:Ergo0}, and we omit their descriptions here.

\subsubsection{Extremal rotating objects: For $a/m=1$}
\label{sec:characterS3}

Configurations satisfying the condition $a/m=1$ are objects in extremal rotation. Moreover, such a condition implies that the extremal mass $m_c/r_0$ and the critical mass $m_h/r_0$, cf.~Eqs.~\eqref{eq:CritMass}, \eqref{eq:critmass2}, and \eqref{eq:masshor}, take the same unit value $m_c/r_0=m_h/r_0=1$. This case is conveniently separated in three disjoint classes of objects.\\

\noindent
{\it (i) For small masses, $0<m/r_0<m_c/r_0$}: 

The configurations with masses in the interval $0<m/r_0<m_c/r_0$ present no horizons, corresponding to rotating regular stars where the matching is done on a timelike surface. The properties of the solutions in respect to the ergosurfaces are basically the same as in the nonextremal rotating cases with $0<a/m<1$, see panels (a) and (b) of Fig.~\ref{fig:Ergo1}.\\ 

\noindent
{\it (ii) For the extremal mass, $m/r_0=m_c/r_0$}: 

In the case where the total mass equals the extremal mass, which also means  $m/r_0=m_{h}/r_0$ and $m/r_0=1$, the matching is done on a lightlike surface located at the extremal (double) horizon, $r_0=r_-=r_+$, and the solution corresponds to an extremal rotating regular black hole. This situation is shown by the curves with label $m/r_0=1.000$ in panels (c) of Figs.~\ref{fig:Deltarxa} and~\ref{fig:gtt}, and the corresponding ergosurfaces and horizons are shown in panel (c) of Fig.~\ref{fig:Ergo1}.\\  

\noindent
{\it (iii) For large masses, $m/r_0>m_c/r_0$}: 

In cases with large masses, i.e., for masses larger than the extremal mass, $m/r_0>m_c/r_0$, which also means $m/r_0>m_{h}/r_0$ and $m/r_0>1$, the matching is done on a timelike surface located at a radius smaller than the extremal horizon, $r_0<r_-=r_+$, and the solutions are rotating regular black holes.
A typical example of this kind of configurations is shown by the curves with label $m/r_0=1.200$ in panels (c) of Figs.~\ref{fig:Deltarxa} and~\ref{fig:gtt}, and the corresponding ergosurfaces and horizons are shown in panel (d) of Fig.~\ref{fig:Ergo1}.

\subsubsection{Overspinning objects: For $a/m > 1$}

In this regime of overspinning solutions no horizons are formed, the matching is done on a timelike surface and all configurations correspond to rotating  regular stars, see the bottom panels of Fig.~\ref{fig:Ergo1}. The properties of the solutions in respect to the ergosurfaces are basically the same as for smaller rotation parameters in the cases with no horizons. The main difference is the presence of two disjoint ergoregions.  As seen from the figure, once the ergosurfaces are formed, two disjoint ergoregions appear and hold for all values of the mass and rotation parameters. Obejct with relatively small masses present the two ergoregions located inside the matter region, while for objects with relatively large masses the ergoregions extend beyond the boundary surface of the matter distribution. Moreover, as the ratio $a/m$ grows the ergoregions become more and more similar to the case of the Kerr geometry.

\section{Conclusion}\label{Sec:Conclusion}

Models for rotating compact objects that may be smoothly matched to the exterior Kerr solution are obtained by following the Gürses-Gursey approach and by using a suitable static spherically symmetric metric as the starting point.
Such a static metric is obtained by taking the same horizon function $g_{rr}= 1/(1 - r^2/R^2 +r^4/A^4)$ as the type VII exact solution by Tolman~\cite{Tolman:1939}, representing a deformed de-Sitter type metric potential with the energy density satisfying a quadratic profile. 
In the present work the deformation on the de Sitter metric is also through the inclusion of a non-isotropic fluid, with the tangential pressures being different from the radial pressure. 
The resulting models bear three free parameters, the mass $m$, the rotation parameter $a$, and a third parameter $r_0$ that carries dimension of length and corresponds to the radius of the boundary surface of the matter distribution in Boyer-Lindquist coordinates.  Among the possible solutions we find rotating regular black holes and other configurations representing rotating regular star-type configurations. By regular solutions we mean spacetimes without curvature singularities. The energy-momentum tensor associated with these solutions is represented by an anisotropic fluid that violates the weak energy condition.

The analysis shows that there is a lower bound on the value of the normalized mass parameter $m_c/r_0$ for spacetimes to present horizons, i.e., spacetimes with mass parameter $m/r_0$ such that  $m/r_0<m_{c}/r_0$ present no horizons for any value of the rotation parameter $a$, and these solutions are interpreted as rotating regular stars. 
For values of mass above the extremal mass $m/r_0> m_{c}/r_0$ and with $a/m<1$, the corresponding geometries exhibit two horizons, representing rotating regular black holes. 
Since the radii of the horizons increase with the mass parameter faster than the matching surface radius, one has that the boundary surface may be timelike, lightlike, or spacelike. 
It is timelike in three situations. The first one is for relatively low masses when no horizons are present and the solutions are rotating regular stars. The second is for intermediate masses when two horizons are present, the solutions are rotating regular black holes and the matching occurs outside the event horizon. The third situation is for very large masses when there are two horizons, the solutions are also rotating regular black holes and the matching occurs inside the Cauchy horizon. It is lightlike when the solutions are rotating regular black holes and the matching surface coincides with one of the horizons. It is spacelike when the solutions are rotating regular black holes and the matching occurs between the two horizons. 

In the present work we focused on finding models for rotating objects and studying the main properties of such solutions, with chief interest on rotating regular black holes. The next step is to study the maximal analytic extension of the rotating regular blackhole configurations displayed above. A further important study is to investigate the stability of the relevant objects found in the present work.

\section*{Acknowledgments}
A. D. M. was financed in part by Coordenação de Aperfeiçoamento
de Pessoal de Nível Superior (CAPES), Brazil, Finance Code 001.
V. T. Z. thanks CAPES, Brazil, Grant No. 88887.310351/2018-00, and Conselho Nacional de Desenvolvimento Científico e Tecnológico (CNPq), Brazil, Grant No. 311726/2022-4.

\section*{Data availability}
Data sharing not applicable to this article as no datasets were generated or analysed during the current study.

\appendix 
\section{The matching conditions on a rotating lightlike surface}\label{appendixA}

To deal with the junction conditions in the particular case where the matching surface coincides with a horizon, we  follow the strategy of Ref.~\cite{Beltracchi2021} in which the authors adapted the Darmois-Israel formalism \cite{Israel:1966rt} so that it can be applied to a lightlike surface.
In this way, the junction conditions on the horizon can be analyzed without following the strategy defined in Ref.~\cite{Barrabes-Israel}, that makes use of the transverse extrinsic curvature constructed from a transverse lightlike normal vector.

In the present case,
the matching parameter is the radial Boyer-Lindquist coordinate $r$. More specifically, the matching surface is a hypersurface defined by $S: r=r_0=\,$constant. 
Thus, the normal vector to $S$ is given by $n^{\mu}=\delta_{r}^{\mu}\sqrt{N\Delta/\Sigma}$, with
$N$ being the norm of $n^{\mu}$ which is chosen to be the lapse function, 
\begin{equation}\label{eq:nomN}
    N\equiv n_{\mu}n^{\mu}=\dfrac{\Sigma\Delta}{\Upsilon},
\end{equation}
where
\begin{equation}\label{eq:upsilon}
    \Upsilon= (r^{2}+a^{2})^{2}-\Delta a^{2}\sin^{2}\theta.
 \end{equation} 
Notice that, differently from the Darmois-Israel formalism, here the norm $N$ is not unity, $N\neq \pm 1$. Since the functions $\Upsilon$ and $\Sigma$ are both positive in the region of interest ($r>0$), the norm $N$ is positive (negative) according to $\Delta(r)$ being positive (negative), and it vanishes as a horizon ($\Delta\to 0$) is approached.
As we shall verify explicitly below, the use of this non-unit normal vector allows us to apply the junction conditions even on a lightlike surface.   

The induced metric on the surface $S$ is 
\begin{equation}\label{eq:metricS}
    ds_{S}^{2}=-Ndt^{2}+g_{\phi\phi}\left(d\phi+\dfrac{g_{t\phi}}{g_{\phi\phi}} dt\right)^{2}+g_{\theta\theta}\, d\theta^{2}
\end{equation}
with the intrinsic coordinates being $\xi^{a}=(t,\,\theta,\,\phi)$. The interior and exterior metrics are given by line element~\eqref{eq:RST} with the mass function $M(r)$ defined by~\eqref{eq:mfunction} for the for interior spacetime region, and by~\eqref{eq:Kerrmass} for the exterior region.

Let us assume initially that the matching surface $S$ is a timelike (or a spacelike)  boundary surface, which means $N^{(\pm)}(r_0)>0$ (or  $N^{(\pm)}(r_0)<0$) and, consequently,  $\Delta^{(\pm)}(r_0)>0$ (or $\Delta^{(\pm)}(r_0)<0$). The continuity of the first fundamental form at the boundary implies in  
$N^{+}=N^{-}$, $g_{\theta\theta}^{+}=g_{\theta\theta}^{-}$, $g_{\phi\phi}^{+}=g_{\phi\phi}^{-}$, and $g_{t\phi}^{+}=g_{t\phi}^{-}$. These conditions are satisfied by the alone relation $M_{in}(r_0) = M_{Kerr}(r_0)$, which, after using Eqs.~\eqref{eq:mfunction} and \eqref{eq:Kerrmass}, gives
\begin{equation}\label{eq:mcond1}
  \dfrac{r_0^3}{R^2}- \frac{r_0^5}{A^4}= 2m.
\end{equation}

The extrinsic curvature components take the form
\begin{equation}
    K_{b}^{a(\pm)}=-g_{(\pm)}^{ad}n_{(\pm)}^{r}\Gamma_{r bd}^{(\pm)},
\end{equation}
with the Latin indexes spanning the intrinsic coordinates on $S$: $a,b=t,\,\theta,\,\phi$. The non identically zero components of $K_{b}^{a(\pm)}$ are
\begin{equation}\label{eq:Ktt}
       K_{t}^{t(\pm)} 
               =-\dfrac{\sqrt{\Upsilon\,}}{2\Sigma}\left(\dfrac{\partial g_{tt}}{\partial r}-\dfrac{g_{t\phi}}{g_{\phi\phi}}\dfrac{\partial g_{t\phi}}{\partial r}\right)\Bigg|^{(\pm)},
\end{equation}
\begin{equation}\label{eq:Kphit}
        K_{\phi}^{t(\pm)} 
      =\dfrac{\sqrt{\Upsilon\,}}{2\Sigma}\left(\dfrac{g_{t\phi}}{g_{\phi\phi}}\dfrac{\partial g_{\phi\phi}}{\partial r}-\dfrac{\partial g_{t\phi}}{\partial r}\right)\Bigg|^{(\pm)},
\end{equation}
\begin{equation}\label{eq:Kthth}
       K_{\theta}^{\theta(\pm)}=- 
        \dfrac{|\Delta|}{2\Sigma\sqrt{\Upsilon\,}}\dfrac{\partial g_{\theta\theta}}{\partial r}\Bigg|^{(\pm)},
\end{equation}
\begin{equation}\label{eq:Kphiphi}
\begin{split}
     K_{\phi}^{\phi(\pm)}&=\dfrac{g_{t\phi}\sqrt{\Upsilon\,}}{2g_{\phi\phi}\Sigma}\left(\dfrac{\partial g_{t\phi}}{\partial r}-\dfrac{g_{t\phi}}{g_{\phi\phi}}\dfrac{\partial g_{\phi\phi}}{\partial r}\right) \\ &\qquad +\dfrac{|\Delta|}{2g_{\phi\phi}\sqrt{\Upsilon\,}}\dfrac{\partial g_{\phi\phi}}{\partial r}\Bigg|^{(\pm)},
\end{split}
\end{equation}
and also the component $K_{t}^{\phi} = g_{t b}\,g^{\phi a} K_{a}^{b} $ that was not written.

The above expressions can be further simplified by noticing that the partial derivatives of $g_{t\phi}$ and $g_{\phi\phi}$ with respect to $r$ may be written in terms of $\partial g_{tt}/\partial r$ as follows,
\begin{equation}
    \dfrac{\partial g_{t\phi}}{\partial r}=-2a\sin^{2}{\theta}\dfrac{\partial g_{tt}}{\partial r},
\end{equation}
\begin{equation}
    \dfrac{\partial g_{\phi\phi}}{\partial r}=2r\sin^{2}{\theta}+2a^{2}\sin^{4}{\theta}\dfrac{\partial g_{tt}}{\partial r}, 
\end{equation}
with $\partial g_{tt}/{\partial r}$ given by
\begin{equation}
    \dfrac{\partial g_{tt}}{\partial r}=\dfrac{2r}{\Sigma}\dfrac{d M(r)}{dr}+\dfrac{2M(r)}{\Sigma}-\dfrac{4r^{2}M(r)}{\Sigma^{2}}.
\end{equation}
Therefore, the smooth junction conditions of the extrinsic curvature on the surface $S$, $K_{b}^{a(-)}=  K_{b}^{a(+)}$, reduce to only one condition, namely, $\partial g_{tt}/{\partial r}\big|^{(-)}= \partial g_{tt}/{\partial r}\big|^{(+)}$, what depends on the mass function $M(r)$ given in Eq.~\eqref{eq:massarot} alone. This condition implies in 
\begin{align}
   & M_{in}(r_0) = M_{Kerr}=m,\label{eq:mconstr1}\\
  &   \dfrac{dM_{in}}{dr}\Bigg|_{r_0}= \dfrac{dM_{Kerr}}{dr}\Bigg|_{r_0} =0. \label{eq:mconstr2}
\end{align}
The first constraint \eqref{eq:mconstr1} implies in relation~\eqref{eq:mcond1}, while the constraint~\eqref{eq:mconstr2} leads to 
\begin{equation}\label{eq:mcond2}
    \dfrac{5r_0^2}{R^2}- \dfrac{3}{A^{4}}=0.
\end{equation}
By combining Eqs.~\eqref{eq:mcond1} and \eqref{eq:mcond2} it follows the two relations shown in Eq.~\eqref{eq:mconstr}. Additionally, by comparing \eqref{eq:mcond2} with Eq.~\eqref{eq:density}, see also Eq.~\eqref{eq:RotEnDen}, it is seen that the boundary surface coincides with the surface of vanishing energy density, which is the surface of zero radial pressure i.e., $\rho(r_0) =0=p_r(r_0)$.

Now it is a straightforward task to verify that all the above functions are well defined in the limit $\Delta\to 0$.
In fact, in such a limit the induced metric on $S$ is given by Eq.~\eqref{eq:metricS} with $N=0$, and is well defined everywhere on such a surface. Then, the continuity of the first fundamental form at the horizon $r=r_0=r_-$, or at $r=r_0=r_-r_+$, is fully satisfied by the constraint~\eqref{eq:mcond1}.
Additionally,  the components of the extrinsic curvature given by Egs.~\eqref{eq:Ktt}--\eqref{eq:Kphiphi} are also well defined with only the component $ K_{\theta}^{\theta(\pm)}$ vanishing in the limit $\Delta\to 0$. 
 Then, the above analysis holds also in such a limit, and the continuity of the second fundamental form is fully satisfied by the conditions \eqref{eq:mcond1} and \eqref{eq:mcond2}, meaning that the smooth match is well behaved even when the matching surface coincides with one of the horizons.

\end{document}